# Evidence of Dirac insulator properties in poly 1,5-dihydro-1,5-diazocine diazene

R. Kevorkyants, V. Ligatchev, and P. Wu

Institute of High Performance Computing (IHPC), 1 Fusionopolis, Way, #16-16 Connexis

Singapore 138632, Singapore

**Abstract**

Existence of heteroaromatic graphene analog is predicted based upon periodic first principles density functional theory calculations. The new material, poly 1,5-dihydro 1,5-diazocine diazene, is a monolayered aromatic (planar) cross-linked polymer with cohesion energy - 6.03 eV/atom. Calculations reveal its *Dirac insulator* properties with narrow (~ 0.057 eV) and nearly direct band gap in close vicinity of Γ-point of Brillouin Zone. The predicted Fermi velocity of charge carriers ranges from ~$3.41 \times 10^5$ m/s to ~$1.63 \times 10^6$ m/s; thus effective mass of those could be up to ~2260 times lower than free electron one. These make proposed material a good alternative to graphene.

PACS: 61.48.Gh; 61.41.+e; 73.22.Pr

Recently made two-dimensional form of crystalline carbon (graphene) opened new horizons for both fundamental science and electronic technology [1]. The unique electronic properties of graphene predominantly originate from unusual behavior of 'gapless excitations' of the 'mass-less Dirac fermions'; e.g. those exhibit extremely high (~$10^6$ m/s) room-temperature Fermi velocity $v_F$ and mobility [~5000 - 15000 cm$^2$/(V*s)] [2]. Therefore, this material seems to be very attractive for electronics, but the zero band gap width $E_G$ prohibits it from being used straightforwardly as a substitute for silicon and/or other semiconductors.

In order to turn graphene to be a *Dirac insulator* (semiconductor), its band gap should be opened, but with preserving the archetypal linear dispersion $E(k)$ of key electron bands. This can be achieved in several ways, for instance, bending in single graphene sheet, implementation of structures comprising of two (doped or twisted) [3, 4] or multiple sheets of graphene [5]; as well as the use of electric field effect [6] can increase the band gap $E_G$ of graphene bilayer up to 0.3 eV. Usually the linear dispersion is not observed in such systems when $E_G > 0.15$ eV [7].

Actually 'Dirac materials' (i.e. those with electronic properties originated from relativistic Dirac fermions) were predicted theoretically long time before the graphene discovery, see Refs. [8 - 11] and references therein. For instance, the majority holes in *graphite* are predicted to be 2D Dirac fermions, which co-exist with normal (massive) 3D majority electrons and 2D minority holes [8]. However, widespread experimental search for such thrilling charge carrier behavior in real materials had been triggered in fact by the graphene triumph. As a result, Dirac fermions were found in pure Bi crystal when placed in a strong (up to 31 Tesla) magnetic field at cryogenic temperatures [10]. Moreover, recently, features of 'topological Dirac insulator' in electrons system of bulk $Bi_{0.9}Sb_{0.1}$ single crystals have been observed in experimental studies of quantum spin Hall effect [9]. This metallic alloy exhibits $E_G \cong 50$ meV, and $v_F$ in the range 7.6x$10^4$ to 1.0x$10^6$ m/s [9]; i.e. in spite of open bandgap, $v_F$ in $Bi_{0.9}Sb_{0.1}$ could be as high as one in the graphene! A *tunable* 'topological Dirac insulator' (with unusual properties of helical Dirac fermions secured up to the temperature $T = 300$ K) have also been reported for $Bi_2Se_3M_x$ ($M_x$ indicates surface doping with Ca or gating control) [11]. Nevertheless, in the present study we will limit ourselves by searching for two-dimensional (2D) systems with electronic properties similar to those of graphene. In particular, the following criteria for the 'graphene analog' (GA) will be used in this study:

1) Its electronic structure is supposed to exhibit linear $E(k)$ dispersion(s) of the highest valence and lowest conduction bands (VB and CB, respectively) simultaneously with an open (non-zero) gap between them;

2) Its CB and VB are expected to be formed by aromatic (i.e. obeying 4n+2 Hückel's rule [12]) conjugated π-electron system, which should facilitate formation of a periodic planar structure;

In addition to these criteria GA must be structurally stable material. Below we give few examples of possible GA candidates.

Probably the most trivial way to create GA with hexagonal topology is to substitute carbon atoms in graphene structure by isoelectronic ones, i.e. atoms of the fourth group of periodic table, Si, Ge, Sn and Pb. It has been reported lately, that, in fact, theory predicts an existence of the stable low-buckled 1D and 2D Ge and Si systems [13]. Furthermore, electronic properties of such systems in planar and low-buckled configurations are very close to those of graphene with similar parameters of Dirac fermions [13]. So far, however, there are no reports on existence of such materials due to significant experimental difficulties in their synthesis.

Substitution of certain carbon atoms by nitrogen ones leads to formation of single-layered 'carbon nitride graphenes' (CNG's). However, their band gap widths range from 2.93 to 5.16 eV, which prohibits appearance of the Dirac fermions [14].

Also known are the monolayered systems of hexagonal symmetry with the general formula $B_xC_yN_z$ ($BC_2N$, $BC_3$, BN [15], B, $AlB_2$, and many others). In the case of BN, for instance, VB and CB are formed by conjugated π-electrons. Its band gap, however, is relatively wide ($E_G \cong 4.5$ eV) at normal conditions (i.e. atmospheric pressure and temperature $T = 273.15$ K), and this material does not possess Dirac fermions as well [15].

An addition of one hydrogen atom to each carbon atom of graphene gives a hydrocarbon called 'graphane', $(CH)_n$. Its direct band gaps of 3.5 eV and 3.7 eV for the 'chair' and 'boat' conformers (respectively), were derived from simulations [16], and are too high for GA. Despite the structural simplicity, graphanes existence was theoretically predicted only relatively recently – in 2007 [16] - and the successful synthesis was carried out in 2009 [17]. In contrast, the evidence for existence of fluorinated 'graphane', $(CF)_n$, dates back to 1934 [18]. Structurally similar to graphene are the monolayers of black phosphorus with the $E_G \cong 1.8$ eV [19]. All three systems are semiconductors with hexagonal quasiplanar structures though lacking Dirac electrons.

In contrast, nearly mass-less Dirac fermions have been predicted theoretically for organic α-(BEDT-TTF)$_2$I$_3$ conductor {BEDT-TTF stands for bis(ethylenedithio) tetrathiafulvalene} [20 - 22]. This material behaves like a metal under atmospheric pressure and $T > 135$ K, while below this transition temperature it turns to be a semiconductor with a resistivity of up to $3*10^3$ Ohm*cm [21]. Such metal-insulator transition is, however, suppressed under a hydrostatic pressure above 1.5 GPa [20, 21]. More comprehensive studies revealed that under such conditions the α-(BEDT-TTF)$_2$I$_3$ material becomes not a real metal, but a semiconductor with an extremely narrow band gap ($E_G < 1$ meV) (see Ref. [20] and references therein). Thus, unusually high mobility [up to $10^5$ cm$^2$/(Vs)] of the charge carriers has been attributed rather to the unique properties of the Dirac fermions with highly anisotropic (Weyl type) dispersion cones at incommensurate 'contact points' with average Fermi velocity $v_F \cong 10^5$ m/s and anisotropy of that close to 10 [22].

Our second GA criterion also holds in the case of α-(BEDT-TTF)$_2$I$_3$: one or both TTF rings become aromatic due to BEDT-TTF oxidation. Indeed, recent DFT study shown that the structure of BEDT-TTF$^+$ is more planar as compared to neutral form because of greater extent of π-conjugation [23]. Additionally, though the (BEDT-TTF)$_2$I$_3$ forms 3D molecular crystal, the ratio of DC conductivities in BEDT-TTF molecules planes to that out-of-plane is about of ~$10^3$, which presumes almost independent 2D molecular layers and electronic sub-systems in such crystals [20].

To construct a 'true' 2D GA, regular planar topology (triangular, square or hexagonal) can be employed. In particular, based on the square topology, we may construct the system shown on the Fig. 1. In this case the 1,5-diazocines 'rings' (placed in vertices of the original regular square topology) are π-conjugated and connected via doubly bonded nitrogen bridges; thus the whole system is π-conjugated.

The geometries of isolated 1,5-dihydro-1,5-diazocine ring, and 2D periodic system were optimized using DFT (PBE) with 6-31G* basis set employing GAUSSIAN 09 program [24]. In the case of geometry optimization of 2D system periodic boundary conditions (PBC) were employed. The band structure of 2D system is calculated using VASP 4.6 program [25]. For band structure calculation charge density was generated using gamma ($\Gamma$) point centered automatic grid of 21x21x1 with vacuum layer thickness of 15 Å. PBE DFT functional together with PAW basis set were employed. The band structure was calculated at 1500 k-points along each high symmetry direction of simple monoclinic cell (*A*[1/2,1/2,0], *B*[-1/2,0,0], *Y*[0,1/2,0]). An additional band structure calculation has been performed over the region surrounding VB maximum and CB minimum ([-0.85:0.85;-0.85:0.85;0]; 2601 points).

The optimized geometry of the structural unit of 2D periodic system is represented in Fig. 1. Formation (cohesive) energy of this structure has been evaluated using standard (for crystalline structures) formula and found to be - 6.03 eV/atom; that ensures structural stability of the 2D system. Usually phonon frequency calculations would be required in order to prove that our system is a planar structure in its ground state. However, in our case, alternative arguments speak in favor of planarity. Although no reference to 1,5 dizocine could be found, the 1,4 diazocine, which is isomeric to our structural unit, does exist and is known to be planar [26]. Thus, we expect 1,5 dizocine molecule to be planar as well. Moreover, aromatic diazene compounds are also known to have planar geometry, e.g. diphenyl diazene. Therefore poly 1,5-dihydro-1,5-diazocine diazene should be planar too.

Features of electronic band structure (in particular, for lowest CB and highest VB) for the proposed GA are illustrated in Figs.2(a), (b). Our calculations reveal 16 extremal points in *A* directions of the BZ, but the one nearest to BZ origin ($\Gamma$ point) corresponds to smallest $E_G$ value and is analyzed thoroughly. This 'first principles' band structure exhibits a certain similarity to that of α-(BEDT-TTF)$_2$I$_3$, obtained within framework of an 'extended Hubbard model' [22]. In particular, two pairs of highly anisotropic VB and CB cones corresponding to linear $E(k)$ dispersion could be seen in vicinity of $\Gamma$ point both for our system [Figs.2(a), (b)] and for α-(BEDT-TTF)$_2$I$_3$ (see Figs. 6, 7 in Ref. [22]). However, our system exhibit the open gap $E_G \cong 0.057$ eV (at $ka_{VB}^{max} = 0.4614$, $ka_{CB}^{min} = 0.5052$ in *A* direction, see Fig.2(b)), while the bandgap in α-(BEDT-TTF)$_2$I$_3$ is closed at an uniaxial pressure $P_a$ exceeding 4.3 kbar. Furthermore, the coordinates of 'contact points' of the VB and CB cones in the latter system are affected by the uniaxial pressure, and move towards the Γ point at the pressure increment; eventually these two contact points vanish in the Γ point at $P_a$ exceeding 40 kbar [22]. This yields finite-mass Dirac fermions in α-(BEDT-TTF)$_2$I$_3$ above 40 kbar. In graphene such contact points are the K points of BZ (K = [0, 2π/3$a$], where $a$ = 2.461 Å is the graphene lattice constant [27]), where symmetric Dirac's cones endow with infinitesimally small relative effective mass $m^*$ of the charge carriers, which eventually provides very high room-temperature Fermi velocity ($v_F$ ~10$^6$ m/s) and mobility [~5000 - 15000 cm$^2$/(V*s)] of the electrons and holes [2].

Slopes of the linear parts of $E(k)$ dependencies in the Fig.2(b) insertion are in the range from ~0.32 eV to ~1.51 eV, which gives $v_F$ varying in the range from ~3.41x10$^5$ m/s to ~1.63x10$^6$ m/s, i.e. it is pretty comparable to that of graphene ($v_F \cong 10^6$ m/s) [6] but with the anisotropy close to 4.8. Similar to α-(BEDT-TTF)$_2$I$_3$ system (where the $v_F$ anisotropy of up to 10 is reported [22]), aforementioned variations in the linear slope of $E(k)$ dependencies and the corresponding speed of the charge carriers are obviously originate from non-symmetric (Weyl type) VB and CB cones in the vicinity of extremal points in Figs. 2(a), 2(b).

Due to non-zero $E_G$ width, the effective mass $m^*$ in our GA becomes finite, and could be roughly estimated from the following relation for the hyperbolic $E(k)$ dispersion [28]: $m^* = E_G/(2v^2)$, which yields $m^* \cong 0.00044\, m_0$ for the branches 2 and 4 of $E(k)$ curve revealed in insertion to Fig.2(b) ($m_0 \cong$ 9.109x10$^{-31}$ kg is the free electron mass). For comparison, in bulk indium antimonide (InSb), $E_G \cong$ 0.17 eV, while electron effective mass and mobility are 0.014 $m_0$ and 7.7·10$^4$ cm$^2$/(Vs), respectively [29]. Since estimated $m^*$ quantity is ~31.8 times lower for our GA than for bulk InSb, anticipated mobility of the charge carriers for the GA could be times higher than in this bulk semiconductor. Moreover, even the lowest $v_F$ quantity in our system surpasses the average one of α-(BEDT-TTF)$_2$I$_3$ by more than 3 times; consequently the charge carrier mobility close to 10$^6$ cm$^2$/(Vs) might be expected for the proposed GA.

In summary, our first-principle (DFT) calculations provide evidence that heteroaromatic 'graphene analog' exhibits properties of a *Dirac insulator*. The new material, poly 1,5-dihydro 1,5-diazocine diazene, is a monolayered two-dimensional aromatic (planar) cross-linked polymer. Calculations show cohesion energy of - 6.03 eV/atom for such polymer, which ensures its stability, narrow (~ 0.057 eV) and nearly direct band gap, and linear dispersion in close vicinity of Γ-point of BZ (***A*** direction, $ka_{VB}^{max}$ = 0.4614, $ka_{CB}^{min}$ = 0.5052). Charge carrier effective mass is predicted to be up to 2260 (!) times lower than the free electron one, while speed of the electrons and holes is ranging from ~3.41x10$^5$ m/s to ~1.63x10$^6$ m/s, which is quite comparable to those of graphene. Based on evaluated cohesion energy and electronic properties, proposed material might be a good alternative to the graphene. Synthesis of such cross-linked polymer structure might be a challenging task. We would, however, like to remind that that experimental challenges at pioneering graphene synthesis and studies also seemed unfathomable at the very beginning [2].

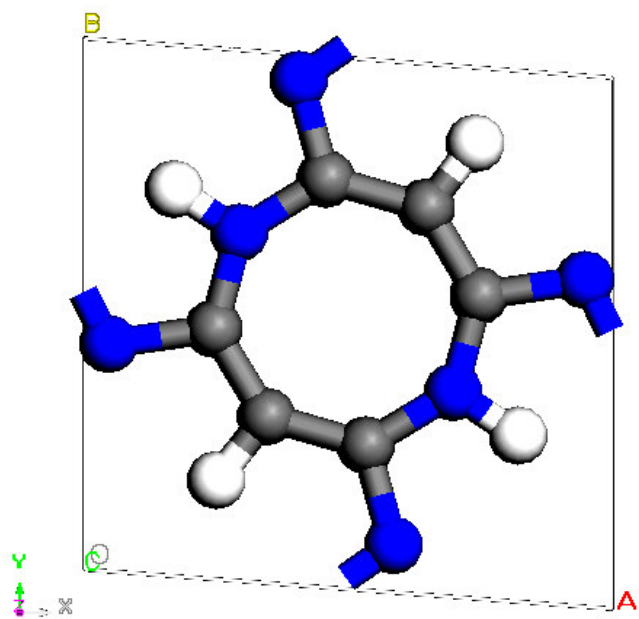

**Figure 1**. (Color online) Ball and stick representation of atomic structure of poly 1,5-dihydro 1,5-diazocine diazene. Carbon, nitrogen and hydrogen atoms are represented by gray, blue (dark gray) and white (light gray) balls, respectively. The rhomb represents the unit cell of 2D periodic (in *OXY* plane) structure. The angle BCA equals to $\cong 93.8^{o}$.

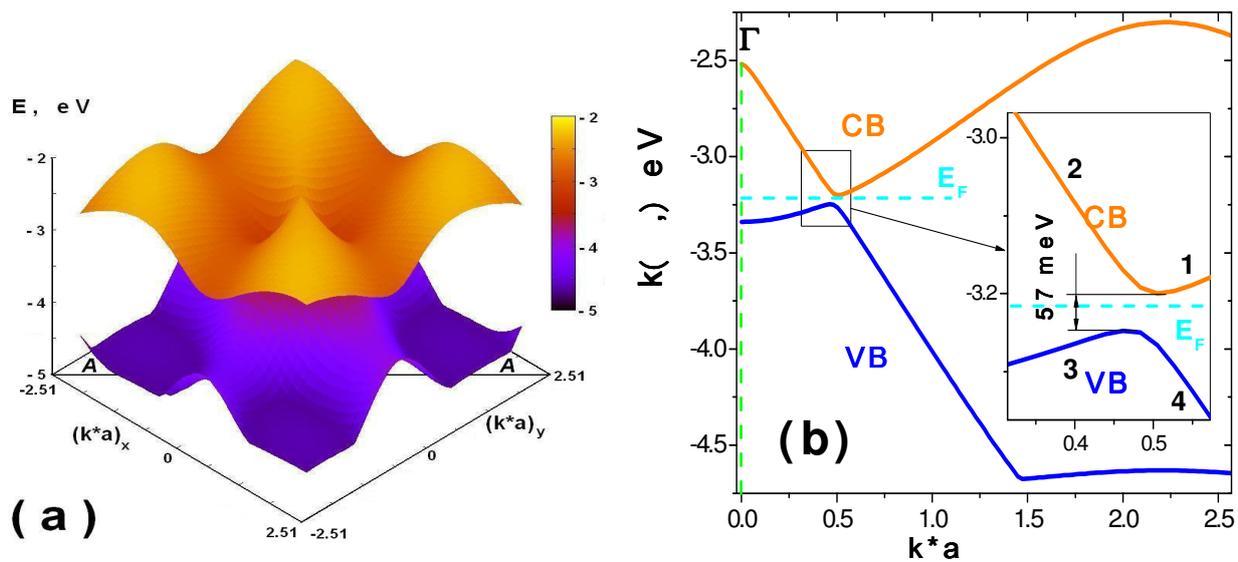

**Figures 2(a), (b)** (Color online). Band diagrams of 1,5-dihydro 1,5-diazocine diazene: (**a**) 3D view of highest VB (bottom) and lowest CB (top); (**b**) cross-section from the figure (**a**) in the diagonal (*A*) direction. Dashed horizontal line in figure (**b**) indicates Fermi level.